 \documentclass[preprint,floats,aps,epsfig,nofootinbib,amssymb]{revtex4}

\usepackage{slashed}
\usepackage{slashed}
\usepackage{graphicx,color}
\usepackage{epsfig}
\usepackage{subfigure}
\usepackage{epsfig}
\usepackage{dcolumn}
\usepackage{bm}
\usepackage{color}

\def\lsim{\mathrel{\rlap{\lower4pt\hbox{\hskip1pt$\sim$}}
    \raise1pt\hbox{$<$}}}         
\def\gsim{\mathrel{\rlap{\lower4pt\hbox{\hskip1pt$\sim$}}
    \raise1pt\hbox{$>$}}}         

\begin{document}

\vspace*{-5.8ex}
\hspace*{\fill}{ACFI-T14-11}

\vspace*{+3.8ex}

\title{Electroweak Baryogenesis, Electric Dipole Moments, and  Higgs Diphoton Decays}

\author{Wei Chao$^{1}$}
\email{chao@physics.umass.edu}

\author{Michael J. Ramsey-Musolf$^{1,2}$ }
\email{mjrm@physics.umass.edu}
 \affiliation{$^1$ Amherst Center for Fundamental Interactions, Department of Physics, University of Massachusetts-Amherst
Amherst, MA 01003\\ 
$^2$California Institute of Technology, Pasadena, CA USA  }

\vspace{3cm}

\begin{abstract}
We study the viability of electroweak baryogenesis in a two Higgs doublet model scenario augmented by vector-like, electroweakly interacting fermions. Considering a limited, but illustrative region of the model parameter space, we obtain the observed cosmic baryon asymmetry while satisfying present constraints from the non-observation of the permanent electric dipole moment (EDM) of the electron and the combined ATLAS and CMS result for the Higgs boson diphoton decay rate. The observation of a non-zero electron EDM in a next generation experiment and/or the observation of an excess (over the Standard Model) of Higgs to diphoton events with the 14 TeV LHC run or a future $e^+e^-$ collider would be consistent with generation of the observed baryon asymmetry in this scenario.


\end{abstract}

\maketitle
\section{Introduction}
The Standard Model of (SM) of particle physics provides an excellent description of a wide variety of laboratory and astrophysical observations.  With the discovery of the Higgs-like scalar at the CERN LHC~\cite{atlas,Aad:2013wqa,cms,Chatrchyan:2013lba}, the SM Higgs mechanism for spontaneous breaking of the gauge symmetry in the SM~\cite{higgs} appears to be a correct description of nature. More precise measurements of Higgs boson properties will help determine whether there exist additional degrees of freedom that participate in electroweak symmetry-breaking or otherwise generate new Higgs boson interactions. Among the most interesting observables is the rate for the Higgs to decay to two photons, a process generated at one-loop order in the SM. At present, the results for this rate obtained from the LHC are somewhat ambiguous. The associated signal  strength, normalized to the SM expectation and measured by the  ATLAS collaboration, is somewhat greater than unity: $\mu_{\gamma\gamma}=1.55^{+0.33}_{-0.28} $~\cite{Aad:2013wqa}, whereas the  CMS collaboration finds  a value completely consistent with the SM:  $\mu_{\gamma\gamma}=0.77\pm0.27$~\cite{Chatrchyan:2013lba} . Combining the $H\to \gamma \gamma, ZZ^*, WW^*$ channels ATLAS obtains the signal strength 
 $\mu=1.33\pm0.14({\rm sat}) \pm0.15({\rm sys})$ for a fixed mass hypothess corresponding to the measured value $m_H=125.5~{\rm GeV}$.   
The corresponding CMS combined result is $\mu=0.80\pm 0.14$ for the fixed mass hypothesis $m_H=125.7~{\rm GeV}$. It is clear that one of the key tasks for the 14 TeV run of the LHC will be to obtain more precise determinations of these rates, as they might show the evidence of Higgs boson couplings to new particles beyond those of the SM.

One motivation for the possible existence of such particles with sub-TeV scale masses is the as yet unexplained origin of the baryon asymmetry of the Universe (BAU). Combining the WMAP seven year results~\cite{Komatsu:2010fb} with those  from other CMB and large scale structure measurements one obtains
\begin{eqnarray}
Y_B\equiv {\rho_B \over s } = (8.82\pm0.23)\times 10^{-11} \; ,
\end{eqnarray}
where $\rho_B$ is the baryon number density, $s$ is the entropy density of the Universe. The recent results obtained by the Planck satellite are consistent, giving  $Y_B = (8.59\pm 0.11)\times 10^{-11}$~\cite{Ade:2013zuv}.

Assuming that the Universe was matter-antimatter symmetric at its birth, it is reasonable to suppose that  interactions involving  elementary particles generated the BAU during the subsequent cosmological evolution. To generate the observed BAU, three Sakharov criteria~\cite{Sakharov:1967dj} must be satisfied in the early Universe: (1) baryon number violation; (2) C and CP violation; (3) a departure from the thermal equilibrium (assuming exact CPT invariance).  These requirements are realizable, though doing so requires physics beyond the SM.  To that end, theorists have proposed a variety of baryogenesis scenarios whose realization spans the breadth of cosmic history. Electroweak baryogenesis (EWBG)~\cite{Morrissey:2012db,Cohen:1993nk,Trodden:1998ym,Riotto:1998bt,Riotto:1999yt,Quiros:1999jp,Dine:2003ax,Cline:2006ts} is one of the most attractive and promising such scenarios, and it is generally the most testable with a combination of searches for new degrees of freedom at the LHC and low-energy tests of CP invariance. In this respect,  searches for permanent electric dipole moments (EDMs) of neutral atoms, molecules and the neutron present provide one of the most powerful probes of possible new electroweak scale CP-violating interactions~\cite{Li:2010ax,Cirigliano:2009yd,Blum:2010by} that may be responsible for EWBG. 

In this paper, we focus on the EWBG and $\mu_{\gamma\gamma}$ implications of the most recent EDM search null result obtained by the ACME experiment using the ThO molecule, from which one may derive a limit on the electron EDM\cite{Baron:2013eja}: $|d_e|<8.7\times 10^{-29} e\cdot {\rm cm}$ with $90\%$ confidence. In earlier work, the authors of Ref.~\cite{McKeen:2012av} studied the interplay of new CPV interactions that may generate both an elementary fermion EDM and a change in the Higgs diphoton rate. One may nominally characterize the impact on the latter through an effective operator ${\tilde c}_h h F{\tilde F}$, where $h$ is the SM Higgs field, $F$ is the electromagnetic field strength with dual ${\tilde F}$, and ${\tilde c}_h$ is a Wilson coefficient of mass dimension $-1$. As shown in that study, the interplay of the two observables may depend sensitively on the particularly ultraviolet completion. In some scenarios, it is possible that the elementary fermion EDM remains rather insensitive to new interactions that may generate a sizable CPV contribution to $\mu_{\gamma\gamma}$, whereas in other cases the EDM limits impose severe constraints on the diphoton decay rate. It is, thus, interesting to ask how this interplay may affect the viability of EWBG, assuming the new interactions provide the requisite ingredients\footnote{For other recent work investigating the interplay of EDMs, Higgs decays, and EWBG, see Ref.~\cite{Shu:2013uua}. In that work, the authors considered a space-time varying CPV phase of the Higgs background field, a complementary approach to the one followed here.}.

Successful EWBG requires a first order electroweak phase transition and sufficiently effective CP violation during the transition. Neither requirement is satisfied in the SM. One simple SM extension of the SM that may allow them to be satisfied is the two Higgs doublet model (2HDM) (for a recent review, see Ref.~\cite{Branco:2011iw})  augmented with vector like fermions (i.e. fermions whose left and right-handed components transform in the same way under the SM gauge group). In this scenario, a strong first order phase transition is induced by the scalar potential (see, {\em e.g.}, Refs.~\cite{Dorsch:2013wja,Dorsch:2014qja}), while  new (physical) CP-violating phases enter the mass matrix of the vector-like fermions as well as the scalar potential. In what follows, we concentrate on the new CP-violation in the Higgs-vector fermion interactions\footnote{For other theoretical and phenomenological implications of vector-like fermions see, for example, 
Refs.~\cite{FileviezPerez:2011pt,Chao:2010mp,Carena:2012xa,Joglekar:2012vc,ArkaniHamed:2012kq,
Almeida:2012bq,Batell:2012mj,Kearney:2012zi,Voloshin:2012tv,Arina:2012aj,Batell:2012zw,Davoudiasl:2012tu,Feng:2013mea,Perez:2013nra,Joglekar:2013zya,Kyae:2013hda,Duerr:2013dza,Schwaller:2013hqa,Huo:2013fga,Dermisek:2013gta,Garg:2013rba,Ishiwata:2013gma,Fairbairn:2013xaa,Dorsner:2014wva,Xiao:2014kba}.}.  

As we show below, these interactions can lead to a resonantly enhanced  CP-violating source for  EWBG. Because the relevant CP-violating parameter space for this relatively simple SM extension is fairly extensive, we restrict our attention to one illustrative parameter space region and demonstrate that the observed BAU can be obtained in this scenario while respecting the electron EDM constraints. A more extensive study of the parameter space will appear in a follow-up study. We also study the impact of the new fermion-scalar interactions on the Higgs diphoton rate. We find the regions favored by the observed Higgs diphoton rate and non-observation of the electron EDM overlap with regions of parameter space wherein a sizable portion of the baryon asymmetry is generated. Looking to the future, we analyze the impact of order-of-magnitude improvements in the sensitivities of both electron EDM and $\mu_{\gamma\gamma}$ probes. For the general case, the electron EDM would provide a substantially more powerful probe of the EWBG-viable parameter space. However, for scenarios where the EDM effect is suppressed ({\em e.g.}, due to mixing with a SM-gauge singlet\cite{McKeen:2012av}), the Higgs diphoton rate may then yield an interesting sensitivity. 

Our discussion of these points in the remainder of the paper is organized as follows: In Section II we describe the model in detail. Section III is devoted to a study of the EDM  and modified Higgs diphoton rate. We study EWBG in Section IV. We summarize in Section VI. A discussion of the re-phasing invariants in this scenario and their relation to the relevant couplings appears in the Appendix.

\section{Model}
We work in the Type-I 2HDM augmented by a pair vector-like fermion doublets $\psi_{L,R}$, transforming as  $(1, 2, -1/2)$ and a pair vector-like fermion singlets $\chi_{L,R}$, transforming as $(1, 1, -1)$. The Yukawa Lagrangian for the new fermions can be written as
\begin{eqnarray}
{\cal L}_{\rm new} = M_\psi  \overline{\psi_L} \psi_R + M_\chi \overline{\chi_L } \chi_R + y_1\overline{\psi_L} {H}_1 \chi_R + y_2 \overline{\psi_L } H_2 \chi_R+y_1^\prime  \overline {\chi_L} H_1^\dagger \psi_R+ y_2^\prime  \overline {\chi_L} H_2^\dagger \psi_R +{ \rm h.c.}  \label{lagrangian}
\end{eqnarray}
The mass matrix for the charged vector like fermions is then
\begin{eqnarray}
{\cal L}_M =\overline{\left( \matrix{\psi_L&\chi_L }\right)} \left(\matrix{ M_\psi & y_1 v_1 + y_2 v_2 \cr y_1^\prime v_1+ y_2^\prime v_2  & M_\chi } \right)\left(  \matrix{\psi_R \cr \chi_R }\right) + {\rm h.c.} \; , \label{mass}
\end{eqnarray}
where the vacuum expectation values~(VEVs) given by $\langle H_i\rangle =v_i$, ($i=1,2$) and $\sqrt{v_1^2 + v_2^2} =174~{\rm GeV}$. Note that Eq.~(\ref{mass}) contains a physical phase that cannot be rotated away by field redefinitions.  

Although we will not consider explicit CP-violation in the scalar potential in our study, it is nevertheless useful to consider the rephasing invariants that one may construct from the parameters in Eq.~(\ref{lagrangian}) and the scalar potential. To that end, we follow Ref.~\cite{Inoue:2014nva}
that considered the soft $Z_2$-breaking interaction $M_{12}^2 H_1^\dag H_2+\mathrm{h.c.}$ with complex $M_{12}^2$. Additional CP-violation may arise from quartic interactions, such as the $Z_2$-symmetric term $\lambda_5^{}  ( H_1^\dagger H_2^{} )^2 + {\rm h.c.}$. As discussed in Appendix A, the rephasing invariants can be written as $\theta_i \equiv {\rm Arg} (y_i^{} y_i^\prime M_\psi^* M_\chi^*)$ $(i=1,2)$, $\theta_3 \equiv {\rm Arg}(y_1^{} y_2^{\prime} M_\psi^* M_\chi^* M_{12}^2)$, $\theta_4 \equiv {\rm Arg} (y_1^\prime y_2^{} M_\psi^* M_\chi^* M_{12}^{2*})$, $\theta_5 \equiv {\rm Arg} ( y_1 y_2 ^* M_{12}^2 )$ and $\theta_6 \equiv {\rm Arg} (y_1^\prime y_2^{\prime *} M_{12}^2)$. Including additional scalar self-interactions,  term would introduce additional rephasing invariants, such as $\theta_7\equiv{\rm Arg}(\lambda_5 M_{12}^{4*} )$. For a more detailed discussion of the CP-violating phases and rephasing invariants, see Appendix A. In what follows, we will assume that the parameters in the scalar potential are all real and concentrate on the effects of CP-violation in the Yukawa sector (\ref{lagrangian}).  

To solve for the mass eigenvalues, we diagonalize the mass matrix by  $2\times 2$ unitary matrices: $U_L^\dagger M U_R={\rm diag }\{\hat m_\psi, \hat m_\chi\}$.  In the mass eigenbasis the mass eigenvalues can be written as
\begin{eqnarray}
\hat m_{\psi,\chi}^2 = {1 \over 2 } \left\{ |M_\psi|^2 + |M_\chi|^2 + {\cal A} + {\cal B} \pm \sqrt{(|M_\psi|^2 -|M_\chi|^2 +{\cal  A}-{\cal B}  )^2 + 4 |{\cal R}|^2  } {\over }\right\} \label{masss}\; .
\end{eqnarray}
where 
\begin{eqnarray}
{\cal A}&=& |y_1|^2 v_1^2 + |y_2|^2 v_2^2 + 2 v_1 v_2 {\rm Re} (y_1 y_2^*)  \; ,\\
{\cal B}&=& |y_1^\prime|^2v_1^2 + |y_2^\prime|^2 v_2^2 + 2 v_1 v_2 {\rm Re}(y_1^\prime y_2^{\prime *}) \; ,\\
{\cal R}&=& M_\psi (y_1^{\prime*} v_1 + y_2^{\prime *} v_2 ) + M_\chi^* (y_1 v_1 + y_2 v_2 ) \; .
\end{eqnarray}
The mixing angles and phases are
\begin{eqnarray}
\theta_L&=&  {1 \over 2 }\arctan \left(  { -2 |{\cal R}| \over |M_\chi|^2 -|M_\psi|^2 + {\cal B}-{\cal A}}\right) \; ,\hspace{1.5cm} \delta_L= -{\rm Arg}({\cal R}) \; , \\
\theta_R&=&  {1 \over 2 }\arctan \left(  { -2 |{\cal Q}| \over |M_\chi|^2 -|M_\psi|^2 - {\cal B}+{\cal A}}\right) \; ,\hspace{1.5cm} \delta_R= -{\rm Arg}({\cal Q}) \; ,
\end{eqnarray}
where ${\cal Q } =M_\psi^*( y_1 v_1 +y_2v_2 ) +M_\chi(y_1^{\prime*} v_1 +y_2^{\prime *}v_2  ) $, $\theta_{L,R}$ and $\delta_{L,R}$ are the mixing angles and phases of $U_{L,R}$,  respectively.  Notice however that $U_{L,R} $ are not completely determined by the following equation,  $ U_L^\dagger MM^\dagger U_L = U_R^\dagger M^\dagger M U_R ={\rm diag}\{\hat {m}^2_\psi,~\hat {m}^2_\chi\}$. They can be multiplied from the right by an arbitrary phase rotation which contains two phases that do not depend on $M$: $U_{L,R} \to U_{L,R} {\rm diag} \{ e^{-i \phi_{\psi_{L,R}}}, ~e^{-i \phi_{\chi_{L,R}}} \}$, where only two combinations,  $\phi_{\psi_L} -\phi_{\psi_R}$ and $\phi_{\chi_L}-\phi_{\chi_R}$, can be solely determined by the parameters in the mass matrix. As shown in Appendix A, $\theta_L$ and $\theta_R $ are separately rephasing invariant, while $\delta_L$ and $\delta_R$ are not. 
  
We note that the mass of the neutral component of $\psi$ is not always below that of the lighter charged state. In order to avoid the existence of a stable charged relic, it is possible to extend the model with additional electroweak singlets $\xi_{L,R}$ whose Yukawa interactions with the 
$\psi_{L,R}$ lead to the presence of a lightest neutral state after electroweak symmetry-breaking. Assuming the new Yukawa couplings are real,  introduction of these new fields and interactions will not affect the Higgs diphoton decay rate, EDM, or EWBG. An extensive analysis of such as scenario will appear in forthcoming work\cite{weiandmichael2}.
For either the latter scenario or for the model considered here when the neutral states are the lightest, one could search for the vector like fermions at the LHC in the diboson plus missing energy channel. As shown, for example, in Ref.~\cite{ArkaniHamed:2012kq}, the present LHC data do not preclude the existence of these fermions for masses in the several hundred GeV and above range.


\section{The Higgs to diphoton rate  and electron EDM}

In the SM, the leading contribution to the Higgs coupling to a diphoton pair is generated by the $W$ boson loop, which is at least four times larger than the next-to-leading contribution from the top quark loop.
New charged fermions generate additional loop level contributions.  The analytical expression for the signal strength $\mu_{\gamma\gamma}$ reads
\begin{eqnarray}
\mu_{\gamma\gamma} ={1 \over s_\beta^2 |A_{\rm SM}^{\gamma\gamma}|^2 } \left\{\left|{2s_\beta\over v } A_1(\tau_W) + {2 N_C Q_t^2\over v s_\beta}  A_{1\over 2 } (\tau_t) + {2{\rm Re }(\eta_i) \over m_i } A_{1\over 2 } (\tau_i)\right|^2  + \left| {2{\rm Im}(\eta_i) \over m_i } \tau_i f(\tau_i )\right|^2 \right\}\label{ggamma}
\end{eqnarray}
with $\Gamma_{\rm SM} (h \to \gamma \gamma ) = \left(  {\alpha^2 m_h^3 \over 1024 \pi^3}\right) |A_{\rm SM}^{\gamma}|^2 $.   Here $s_\beta=\sin\beta$ with $\beta=\arctan(v_1/v_2 )$, $N_C=3$ is the number of the colors, $Q_{t,\chi,\psi}$ are electric charge of the top quark and new fermion  in units  of $|e|$, $\tau_i=4\hat m_i^2 /m_h^2~(i=\psi,\chi)$,  and the $\eta_i$ are couplings of new charged fermions to the SM Higgs boson. For $\tan\beta\sim 1$, global fits to the LHC Higgs boson rates imply that the $H_1^0-H_2^0$ mixing angle $\alpha$ is $-0.875(-0.808)$~\cite{Inoue:2014nva}, for the Type-I(II) 2HDM. For illustrative purposes, we will take $H_1^0$ to be the SM-like Higgs boson with $\cos\alpha=1$. In this limit, the  $\eta_i$ are given by 
\begin{eqnarray}
\eta_\psi  &=&+ { |y_1| \over \sqrt{2} }  c_L s_R  {\rm exp} \left\{ i {\rm Arg} \left[ |y_1|^2 +\left| {y_1 y_2 v_2 \over v_1}\right| e^{i\theta_5} +\left |{M_\chi y_1 y_1^\prime \over M_\psi }\right|e^{i \theta_1 } +\left |{M_\chi y_1 y_2^\prime v_2  \over M_\psi  v_1}\right| e^{i\theta_3} \right]\right\} \nonumber \\
&&+{ |y_1^{\prime }| \over \sqrt{2}} c_R s_L  {\rm exp } \left\{ i {\rm Arg} \left[ |y_1^\prime|^2 +\left| {y_1^\prime y_2^\prime v_2 \over v_1}\right| e^{i\theta_6} +\left |{M_\chi y_1 y_1^\prime \over M_\psi }\right|e^{i \theta_1 } +\left |{M_\chi y_1 y_2^\prime v_2  \over M_\psi  v_1}\right| e^{i\theta_4}  \right]\right\} \label{odd1}\\
\eta_\chi
&=& -  { |y_1 | \over \sqrt{2}}  c_R s_L {\rm exp } \left\{  i {\rm Arg} \left[  |y_1|^2 +\left|{y_1 y_2 v_2 \over v_1} \right|e^{i\theta_5}+ \left| { M_\psi y_1 y_1^\prime \over M_\chi } \right| e^{i \theta_1}+ \left| {M_\psi y_1 y_2^\prime v_2 \over M_\chi v_1}\right|e^{i\theta_3}\right] \right\} \nonumber \\
&&-{ |y_1^\prime | \over \sqrt{2} }  c_L s_R {\rm exp}\left\{ i {\rm Arg} \left[  |y_1^\prime|^2 + \left| {y_1^\prime y_2^\prime v_2 \over v_1}\right| e^{i\theta_6} +\left |{M_\psi y_1 y_1^\prime \over M_\chi }\right|e^{i \theta_1 } +\left |{M_\psi y_1 y_2^\prime v_2  \over M_\chi  v_1}\right| e^{i\theta_4}  \right]\right\} \label{odd2}
\end{eqnarray} 
We refer the read to Appendix A for details of the derivation of Eqs.~(\ref{odd1}) and (\ref{odd2}). The explicit expressions for $A_{1/2} (x)$ and $f(x)$ can be found in Ref. \cite{Carena:2012xa}. In the presence of mixing between the two neutral CP-even scalars,  the RHS of Eq.~ (\ref{ggamma}) is multiplied by a factor of $\cos^2 \alpha $. 

\begin{figure}[h]
\begin{center}
\includegraphics[width=7.5cm,height=7cm,angle=0]{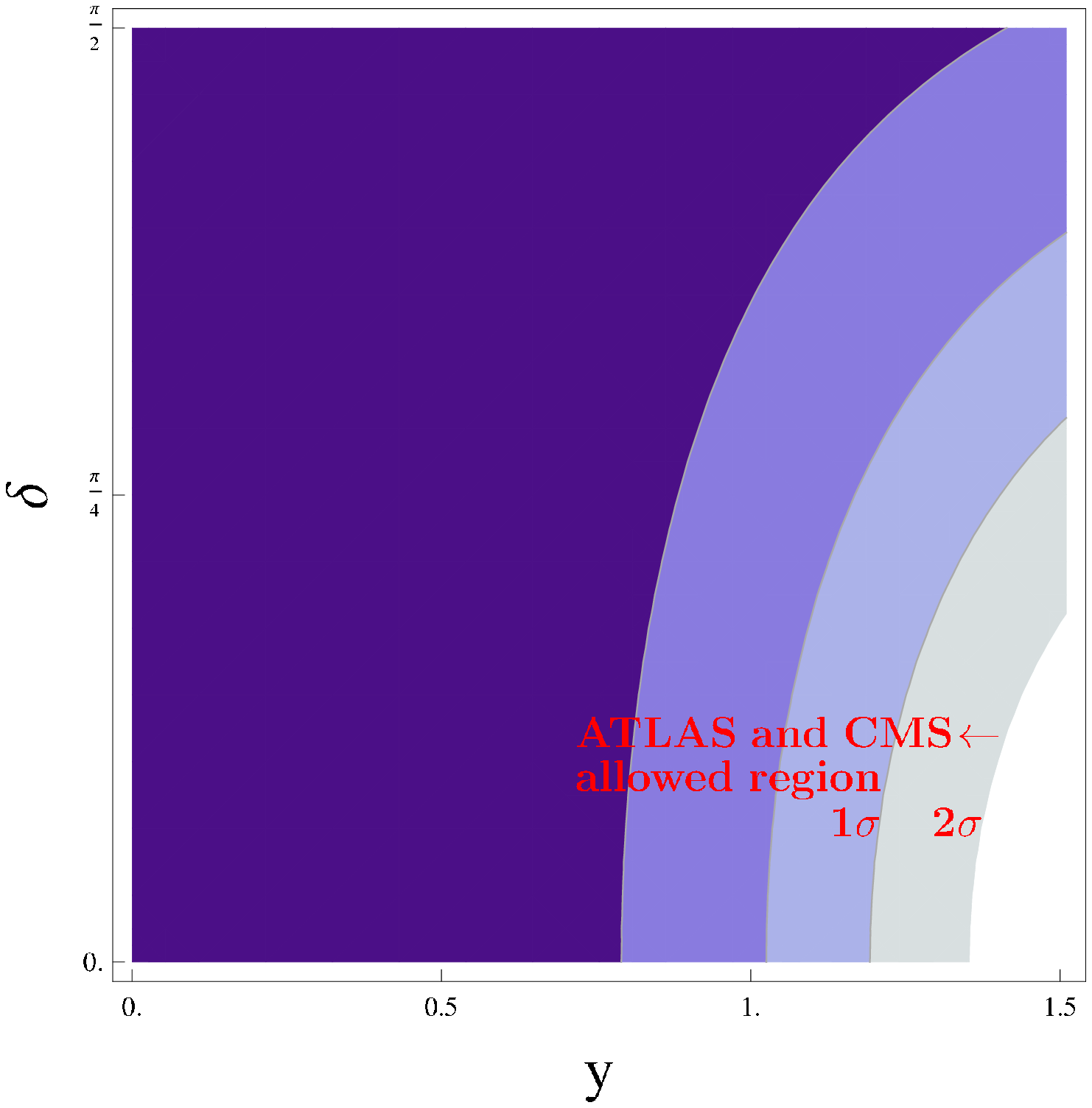}
\hspace{0.5cm}
\includegraphics[width=7.5cm,height=7cm,angle=0]{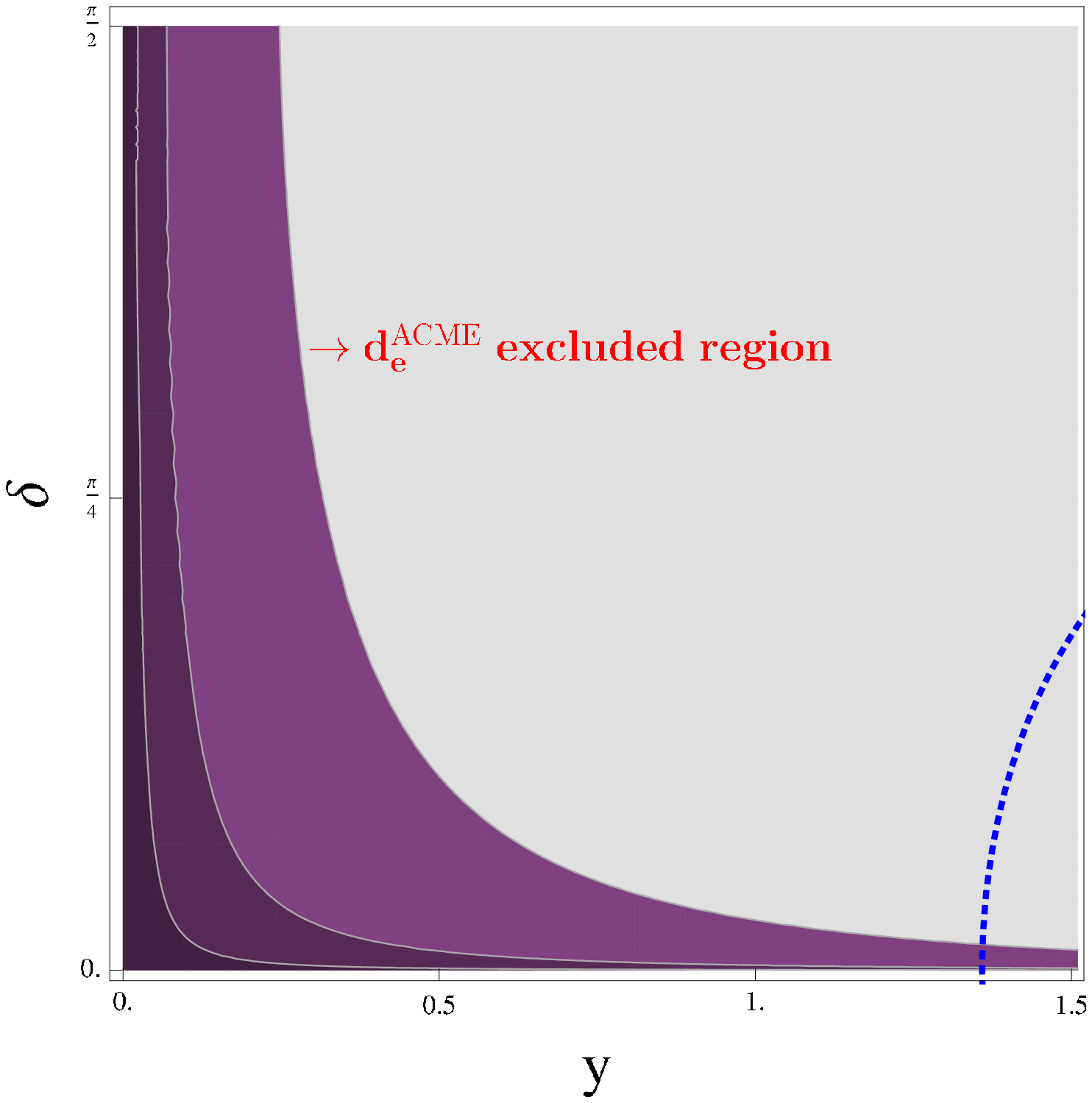}
\end{center}
\caption{The contours of constant $\mu_{\gamma \gamma}$ (left panel) and $d_e\times 10^{29}$ (right panel) in the $y$-$\delta$ plane, where we set $\delta$ to be the phase of $y_1$, $y\equiv |y_1|=y_1^\prime$, $y_2=0.5$, $y_2^\prime =0$, $\tan \beta=10$, $M_\chi=350~{\rm GeV}$ and $M_\psi=380~{\rm GeV}$.  For the contours of  $\mu_{\gamma \gamma}$, we have $\mu_{\gamma \gamma}\in [1,~1.1],~[1.1,~1.2],~[1.2,~1.32],~[1.32,~1.52]$ and $\mu_{\gamma \gamma}>1.52$ from the left to the  right.  For the contours of electron EDM, we have $d_e<1\times 10^{-30}$, $d_e \in [1\times 10^{-30},~1\times 10^{-29}],~[1\times 10^{-29},~1.025\times10^{-28}]$ and $d_e >1.025\times10^{-28}$, from the left to the right.  The grey region in the right panel is excluded by ACME at $95\%$ C.L.. The region to the left of the dashed blue line in the right panel indicates the $\mu_{\gamma\gamma}$ allowed region at 95\% C.L.. }  \label{density}\end{figure}

The CP-odd Yukawa couplings given in Eqs. (\ref{odd1}) and (\ref{odd2}) generate an elementary fermion EDM {\em via} two-loop Barr-Zee diagrams~\cite{Barr:1990vd}. For illustrative purposes, we will work in the limit that the masses of the remaining neutral scalars (one CP-even and one CP-odd) and charged scalars are sufficiently heavy that the dominant contributions arise from exchange of (a) the SM-like neutral scalar and a photon and (b) a $W^+W^-$ pair. CP-violation enters the latter contribution through the relative phase of left- and right-handed charged currents. The result is well-known, and specializing to our case we arrive at the following result for the electron EDM~{\cite{McKeen:2012av}}: 
\begin{eqnarray}
d_e&=& d_e^{(2l)} \sin\Theta  \sin2\theta_L \sin 2\theta_R { \hat m_\chi \hat m_\psi \over m_W^2 } {\alpha_W \over 8 \alpha } \left[  {j(z_1, z_0)\over z_1} -{j(z_2, z_0)\over z_2}\right] \nonumber \\&+&  \sum_{i= \chi, \psi}d_e^{(2l)} Q_i^2 {\rm Im} [\eta_i] {2 v_1\over \hat m_i} g\left(  {\hat m_i^2 \over m_h^2}\right) \; , 
\end{eqnarray}
where $d_e^{(2l)} \approx 2.5\times10^{-27}$ $e\cdot {\rm cm}$, $z_1 = \hat{m}_\chi^2 /m_W^2$, $z_2 = \hat{m}_\psi^2 /m_W^2$, $z_0 = |M_\psi|^2 /m_W^2$,  the loop functions $g(x)$ and $j(x,~y)$ are given in Ref.~{\cite{McKeen:2012av}}. The first term arises from $W^+W^-$-exchange and depends on $\Theta\equiv\delta_L -\delta_R + Arg(M_\psi) -Arg(M_\chi)$,  the relative phase between the left- and right-handed currents that  is rephasing invariant as  shown in  Appendix A. The second term is generated by the $H_1^0\gamma$-exchange graphs. Note that $\Theta\to 0$ in the limit that $|M_\psi| = |M_\chi|$, whereas the second term remains non-zero in this regime. As we discuss in Section IV, EWBG is most viable in the regime that $|M_\psi| \sim |M_\chi|$, in which case $d_e$ will be dominated by the $H_1^0\gamma$-exchange contribution. 

Apparently, Eqs.~(\ref{odd1}) and (\ref{odd2}) are rephasing invariant and can be expressed as  functions of  $\theta_{j}$ $(j=1,3,4,5,6)$. Since we have assumed that the mixing between the neutral Higgs fields is negligible, the phase that gives the dominant contribution to the CP-odd $H\gamma\gamma$ coupling should  be $\theta_1$~\cite{Voloshin:2012tv}.  Thus, the couplings in Eqs.~(\ref{odd1}) and (\ref{odd2}) should govern both the magnitude of any impact on the $H\to\gamma\gamma$ rate as well as $d_e$ in the $|M_\psi| \sim |M_\chi|$ regime that is most relevant for EWBG. Doing so is particularly timely in light of the recent ACME result\cite{Baron:2013eja}, from which an order of magnitude more stringent $d_e$ limit has been obtained (assuming the absence of any other CP-violating sources in the ThO molecule). As we will show in Section IV, the resulting constraints on the nevertheless leave ample room for successful EWBG. 

To illustrate, we will work in a simplified region of parameter space that still allows us to assess general features of the EWBG-EDM-Higgs diphoton interplay. Specifically, we assume $M_\chi$, $M_\psi$ to be real and set $y_1 =y e^{i \delta}$, $y_1^\prime =y $ where $y$ is a real parameter. 
We plot in the left panel of Fig. \ref{density} the contours of constant $\mu_{\gamma \gamma}$ in the $y-\delta$ plane, choosing $\tan \beta=10$, $y_2=0.5$, $y_2^\prime=0$, $M_\chi=350~{\rm GeV}$ and $M_\psi=380~{\rm GeV}$.  Clearly, the impact of this scenario on the Higgs diphoton rate is consistent with the combined ATLAS and CMS $\mu_{\gamma \gamma }$ value {$1.12\pm0.40(2\sigma)$} for a rather wide range of the $y-\delta$ parameter space.  In the right panel of Fig. \ref{density} we plot the contours of constant $d_e\times 10^{29}$ in the $y-\delta$ plane using the same input parameter choices.  The contour line on the rightmost corresponds to the current experimental upper limit on $d_e$ obtained by ACME experiment\cite{Baron:2013eja} . The successive contours to the left of the exclusion line correspond respectively to $d_e$ being one and two orders of magnitude smaller than the current limit.  We observe that the present $d_e$ constraints rule out most of the available parameter space at large $y$ and $\delta$. The diphoton decay rate displays a sensitivity only for relatively small values of the CPV phase, wherein the effect arises largely through the CP-conserving operator $hFF$. This feature is consistent with the general expectations based on the study of Ref.~\cite{McKeen:2012av}.  As we discuss below,  future determinations of $\mu_{\gamma\gamma}$ may retain an interesting sensitivity to the new fermion masses for values of $y$ and $\delta$ giving rise to successful EWBG, thereby complementing the information provided by $d_e$.

\section{Electroweak Baryogenesis}

We now proceed to study EWBG in this scenario.  The three Sakharov conditions are realized in the following way. First, the two Higgs doublets potential can induce a strongly first order electroweak phase transition (EWPT) at temperatures $T\sim 100$ GeV, which provides a departure from equilibrium~\cite{Cline:1996mga,Fromme:2006cm}.  During the EWPT, bubbles of broken electroweak symmetry nucleate  and expand in a background of unbroken symmetry, filling the Universe to complete the phase transition.  Second, the CP-violation arises from the complex phases in the couplings of the new fermions to the Higgs scalars. The phase induces CP-violating interactions at the walls of the expanding bubbles, where the Higgs vacuum expectation value is spacetime dependent, leading to the production of a CP-asymmetric charge density.  This CP-asymmetry diffuses ahead of the advancing bubble and is converted into a net density of left-handed fermions, $n_L$,  through  inelastic interactions in the plasma. Third, baryon number is violated by the sphaleron processes. The presence of nonzero $n_L$ biases the sphaleron processes, resulting in the production of the baryon asymmetry~\cite{Kuzmin:1985mm}. 

We ignore the wall curvature in our analysis so all relevant functions depend on the variable $\bar z = z +v_w t$, where $v_w$ is the wall velocity; $\bar z<0, >0 $ correspond to the unbroken and broken phases, respectively. Working in the closed time path  formulation and under the ``vev-insertion" approximation~\cite{Riotto:1998zb,Chung:2009qs,Blum:2010by,Lee:2004we}, we compute the CP-violating source induced by the Higgs mediated processes $\psi\to \chi\to \psi$,
\begin{eqnarray}
S^\psi_{\rm CP} (x)  &=&{\rm Im} \left\{ |y_1^{ } y_2^{}|e^{i\theta_5}+|y_1^\prime y_2^{\prime }| e^{i\theta_6}\right\} v^2 \dot{\beta}\int {k^2 d k \over \pi^2 \omega_\chi \omega_\psi }{\rm Im } \left\{ (\mathcal{E}_\psi  \mathcal{E}_\chi^* -k^2 ) {n(\mathcal{E}_\psi) -n(\mathcal{E}_\chi^*) \over (\mathcal{E}_\psi - \mathcal{E}_\chi^*)^2 } \right.  \nonumber  \\&& + \left.  (\mathcal{E}_\chi  \mathcal{E}_\psi +k^2 ) { n(\mathcal{E}_\chi) + n(\mathcal{E}_\psi) \over (\mathcal{E}_\chi +\mathcal{E}_\psi)^2 }  \right\} \; ,\label{source1} \\
S_{\rm CP}^{\psi \prime} (x) &=& {\rm Im } \left\{|y_1y_2^\prime|e^{i\theta_{3}} -|y_1^{\prime  } y_2^{}|e^{i\theta_4} \right\} v^2 \dot{\beta}\int {k^2 d k \over \omega_\chi \omega_\psi \pi^2 } |M_\chi| |M_\psi | {\rm Im} \left\{ {n(\varepsilon_\chi ) -n(\varepsilon_\psi^*) \over (\varepsilon_\chi -\varepsilon_\psi^*)^2 } \right. \nonumber \\ &&\left. -{n(\varepsilon_\chi ) + n(\varepsilon_\psi ) \over (\varepsilon_\chi + \varepsilon_\psi)^2 }\right\}  \; . \label{source2}
\end{eqnarray}
where $n(x) =1/exp(x)+1$ is the fermion distribution;  $\varepsilon_{\chi,\psi} =\omega_{\chi,\psi} -i\Gamma_{\chi,\psi}$ are complex poles of the spectral function with $\omega^2_{\chi,\psi}={k^2 +m^2_{\chi,\psi}}$, where $m_{\chi, \psi}$ and $\Gamma_{\chi,\psi}$ are the thermal masses and thermal rates of $\chi$ and $\psi$, respectively.   As can  be seen from eqs. (\ref{aa}$\sim$\ref{dd}) of Appendix A,  $\theta_{i}$ are not independent. As a result, CP-violating phases in Eqs. (\ref{source1},\ref{source2}) can be  correlated with those in Eqs.~(\ref{odd1},\ref{odd2}).  As indicated in Section III, for illustrative purposes we assume $y_1$ contains the only CP phase and $y_2^\prime =0$. In this case, the only non-vanishing phases are $\theta_{1,5}$, implying a non-vanishing $S_{\rm CP}^{\psi} (x) $ but zero $S_{\rm CP}^{\psi \prime} (x) $. For the more general case, both CP-violating sources will contribute to the asymmetry generation. Before proceeding, we note that the vev insertion approximation used in obtaining Eqs.~(\ref{source1},\ref{source2}) is likely to lead to an overly large baryon asymmetry by at least a factor of a few, though a definitive quantitative treatment of the CPV fermion sources remains an open problem. The results quoted here, thus, provide a conservative basis for assessing the EDM and Higgs diphoton restrictions on the EWBG-viable parameter space. For a detailed discussion of the theoretical issues associated with the computation of the CPV source terms, see Ref.~\cite{Morrissey:2012db} and references therein.

We now derive the transport equations that govern the asymmetry generation. In general, these equations depend on the densities of first and second generation left-handed quark doublets, $q_{kL}$, $k=1,2$; first and second generation right-handed quarks, $u_R$, $d_R$, $c_R$, and $s_R$; third generation left-handed quark doublets $Q$ and right-handed singlets, $T$ and $B$; the corresponding lepton densities; that for neutral scalars $H$; and the new fermions $\psi$ and $\chi$.  Since the new fermions have Dirac mass terms in Eq.~(\ref{lagrangian}) it makes sense to consider a single density for the Dirac fermions $\psi$ and $\chi$ constructed from  the $\psi_{L,R}$ and $\chi_{L,R}$, respectively. 

Several physical considerations then allow us to reduce the number of transport equations. Since the SM lepton Yukawa couplings are small compared to those of the third generation quarks, any reaction that converts a non-vanishing $H$ into lepton densities will occur too slowly to have an impact on the dynamics of the plasma ahead of the advancing bubble wall. Consequently, we may omit the SM leptons from the set of transport equations.
Moreover, since all light quarks are mainly produced by strong sphaleron processes and all quarks have similar diffusion constants, baryon number conservation on time scales shorter that the inverse electroweak sphaleron rate implies the approximate constraints $q_{1L} =q_{2L}=-2u_R=-2d_R=-2s_R=-2c_R= -2B =  2( Q+T)$. The resulting set of transport equations can then be written as
\begin{eqnarray}
\partial^\mu Q_\mu &=&+ \Gamma_{m_t} \left( {T\over k_T}  -{Q \over k_Q} \right) + \Gamma_{Y_t} \left( {T\over k_T}-{Q\over k_Q } -{H\over k_H}   \right) + 2\Gamma_{ss} \left( {T \over k_T} -2{Q \over k_Q} + 9 {B \over k_B }  \right)  \; , \label{a}\\
\partial^\mu T_\mu &=&- \Gamma_{m_t} \left( {T\over k_T}  -{Q \over k_Q} \right) - \Gamma_{Y_t} \left( {T\over k_T}-{Q\over k_Q } -{H\over k_H}   \right)  -\Gamma_{ss} \left( {T \over k_T} -2{Q \over k_Q} + 9 {B \over k_B }  \right)  \; ,\label{b}\\
\partial_\mu \psi_{\mu} &=& +\Gamma_\psi^+ \left( {\chi\over k_\chi } + { \psi \over k_\psi} \right) + \Gamma_\psi^- \left( { \chi \over k_\chi} - {\psi \over k_\psi} \right) + \left(\sum_i\Gamma_{y_i} \right)\left( {\chi \over k_\chi} -{H \over k_H} -{\psi \over k_\psi } \right) + S^\psi_{\rm CP}   \; , \label{c}\\
\partial_\mu \chi_{\mu} &=& -\Gamma_\psi^+ \left( {\chi \over k_\chi } + { \psi \over k_\psi} \right) - \Gamma_\psi^- \left( { \chi \over k_\chi} - {\psi \over k_\psi} \right) - \left(\sum_i\Gamma_{y_i} \right) \left( {\chi \over k_\chi} -{H\over k_H} -{\psi \over k_\psi } \right) - S^\psi_{\rm CP}  \; , \\
\partial_\mu H_{\mu } &=&    \Gamma_{Y_t}\left( {T \over k_T } - {H\over k_H } -{Q \over k_Q }  \right) + \left(\sum_i \Gamma_{y_i}\right) \left( {\chi \over k_\chi} -{H \over k_H } - {\psi \over k_\psi}\right)  -\Gamma_h {H\over k_H } \; ,  \label{d}
\end{eqnarray}
where $\partial^\mu =v_w {d \over d \bar z} -D_a {d^2 \over d \bar z^2}$ in the planar bubble wall approximation with $D_a$ being the diffusion constant, while  $n_i$ and $k_i$ are the number density and the statistical factor for particle $``i"$, respectively.
The coefficient $\Gamma_{y_a}$  denote the interaction rates arising from top quark and new fermions;
$\Gamma^{\pm}_i$ and  $\Gamma_{h}$ denote the CP-conserving scattering rates of  particles with the background Higgs field within the bubble; and $\Gamma_{ss} =6\kappa^\prime {8\over 3} \alpha_s^4 T$ is the strong sphaleron rate, where $\alpha_s$ is the strong coupling and $\kappa^\prime\sim{\cal O}(1)$. 

\begin{table}[h]
\centering
\begin{tabular}{cc|cc |cc|cc}
\hline \hline 
$T$ & $100~{\rm GeV}$ & $\Delta \beta$ & $0.015$  & $D_Q$ & $6/T$ &$D_H$ & $100/T$ \\
$v(T)$ & 125~${\rm GeV}$ & $v_w$ & 0.05 & $D_\chi$ &$380/T$  & $M_\psi$ & 250~${\rm GeV}$\\
$L_w $ & $25/T$ &$\tan \beta $ &$15$ &$D_\psi$ & $100/T$ &$M_\chi$ & 250~${\rm GeV}$\\
\hline \hline
\end{tabular}
\caption{ Input parameters at the benchmark point.  }\label{aaa}
\end{table}
The transport coefficient $\Gamma_{\psi}$ can be written as: $\Gamma_\psi=6|y|^2I_F(m_\psi,m_\chi,m_h)/T^2 $, which describes the rate for the processes $\chi\leftrightarrow \psi H$ to occur. We refer the reader to Ref.~\cite{Cirigliano:2006wh} for the general form of $I_F$. The interaction time scale is $\tau_\psi\equiv \Gamma_\psi^{-1}$. In principle, if $\tau_\psi\ll$ the diffusion time\footnote{ $\tau_{\rm diff}$ is the time that it takes for charge, have been created at the bubble wall and having diffused into the unbroken phase, to be recaptured by the advancing bubble wall and be quenched through CP-conserving scattering within the phase of broken electroweak symmetry.} $\tau_{\rm diff}$,  then this interaction occurs rapidly as the charge density diffuses ahead of the advancing wall, leading to chemical equilibrium.  Numerically, we have $\tau_{\rm diff}\sim 10^4/T$~\cite{Chung:2009cb} and $\tau_\psi\sim 10^3/T$ by taking $y_i\sim0.25$, which is { consistent with the $\mu_{\gamma\gamma}$ constraints indicated in Fig. \ref{density}}. In this case, the new fermion Yukawa interaction is in chemical equilibrium, and the relation $ {\chi \over k_\chi} -{H \over k_H} -{\psi \over k_\psi } =0$ is satisfied.  The sum of transport equations for $\psi$ and $\chi$ gives $v_w(\psi+\chi) -(D_\psi \psi^{\prime \prime} + D_\chi \chi^{\prime\prime} ) =0 $, which implies $D_\psi \psi = -D_\chi \chi $ in the static limit~\cite{Chung:2009cb}. Therefore, we have
\begin{eqnarray}
\psi\equiv \tau_\psi H={k_\psi \over k_H} {k_\chi D_\chi \over k_\chi D_\chi + k_\psi D_\psi } H \; ,  \hspace{1cm}
\chi \equiv \tau_\chi H =-{k_\chi \over k_H} {k_\psi D_\psi \over k_\chi D_\chi + k_\psi D_\psi } H\; .
\end{eqnarray}
When top quark Yukawa interaction and strong sphaleron process are in chemical equilibrium, we have 
\begin{eqnarray}
Q\equiv \tau_Q H= {k_Q \over k_H } {k_B-9k_T \over 9k_Q +k_B + 9k_T} H\;, \hspace{1cm }T\equiv \tau_T H= {k_T \over k_H } {9k_Q + 2 k_B \over 9k_Q + k_B + 9k_T } H \; .  
\end{eqnarray}
In terms of $H$, the left-handed fermion charge density becomes $n_L(z) = (5\tau_Q + 4 \tau_T  ) H $.

Since $n_B$ is determined by $n_L$, all that remains is to solve for the Higgs charge density.  The transport equations can be reduced into a single equation for  $H$ by taking the appropriate linear combination of equations:  $(\ref{a}) +2 \times (\ref{b}) +(\ref{c}) +(\ref{d})$. Lastly, the BAU is given by
\begin{eqnarray}
n_B= -{3 \Gamma_{ws} \over 2 D_Q \lambda_{+} } \int^{-L_w/2}_{-\infty} dz n_L(z) e^{-\lambda_{-} z }
\end{eqnarray}
with  $\lambda_{\pm} ={1 \over 2 D_Q} (v_w\pm \sqrt{v_w^2 +4D_Q R}$,  where $R\sim 2\times 10^{-3}~{\rm GeV}$ is the { inverse} washout rate for the electroweak sphaleron transitions.

The computation of $n_B/s$ relies upon many other numerical inputs; our choices are listed in Table. I. The diffusion constants were calculated in Ref~\cite{Joyce:1994bi,Joyce:1994zn} with $D_{\chi} ={380\over T}$ and $D_\psi = {100\over T }$. The fact that $D_\psi \ll D_\chi $ enhances the left-handed  lepton charge. The bubble wall velocity $v_w$, thickness $L_w$, profile parameters $\Delta \beta$ and $v(T)$ describe the dynamics of the expanding bubbles during the EWPT, at the temperature $T$. We take the Higgs profile to be
\begin{eqnarray}
v(z) &\simeq&  {1 \over 2 } v(T) \left\{ 1 + {\rm tanh}\left( 2 \alpha {z \over L_w}\right)\right\} \; ,  \\
\beta(z) &\simeq& \beta_0 (T) - {1 \over 2 } \delta \beta \left\{1 -  {\rm tanh}\left( 2 \alpha {z \over L_w}\right)\right\} \; ,
\end{eqnarray}
following Ref.~\cite{Moreno:1998bq,Carena:2000id,Carena:1997gx}.  The sphaleron rates are $\Gamma_{ws} =6 \kappa \alpha_s^4 T$ and $\Gamma_{ss}=6 \kappa^\prime \alpha_s^4 T {8 \over 3 } $,  where $\kappa_{ws} = 22\pm 2$ and $\kappa_{ss} ={\cal O }(1)$.  

\begin{figure}[h]
\begin{center}
\includegraphics[width=7.0cm,height=6cm,angle=0]{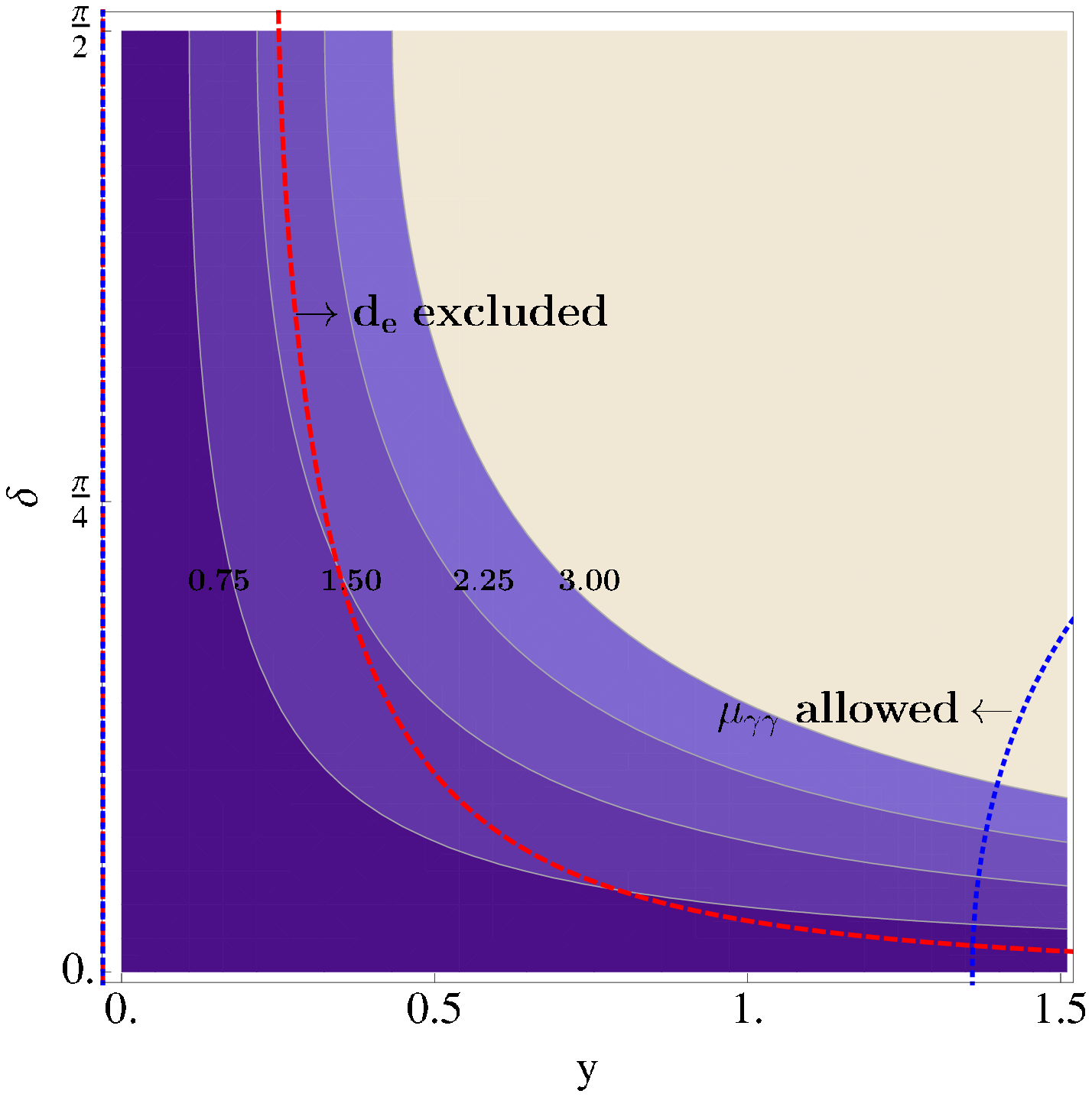}
\hspace{0.5cm}
\includegraphics[width=7.0cm,height=6cm,angle=0]{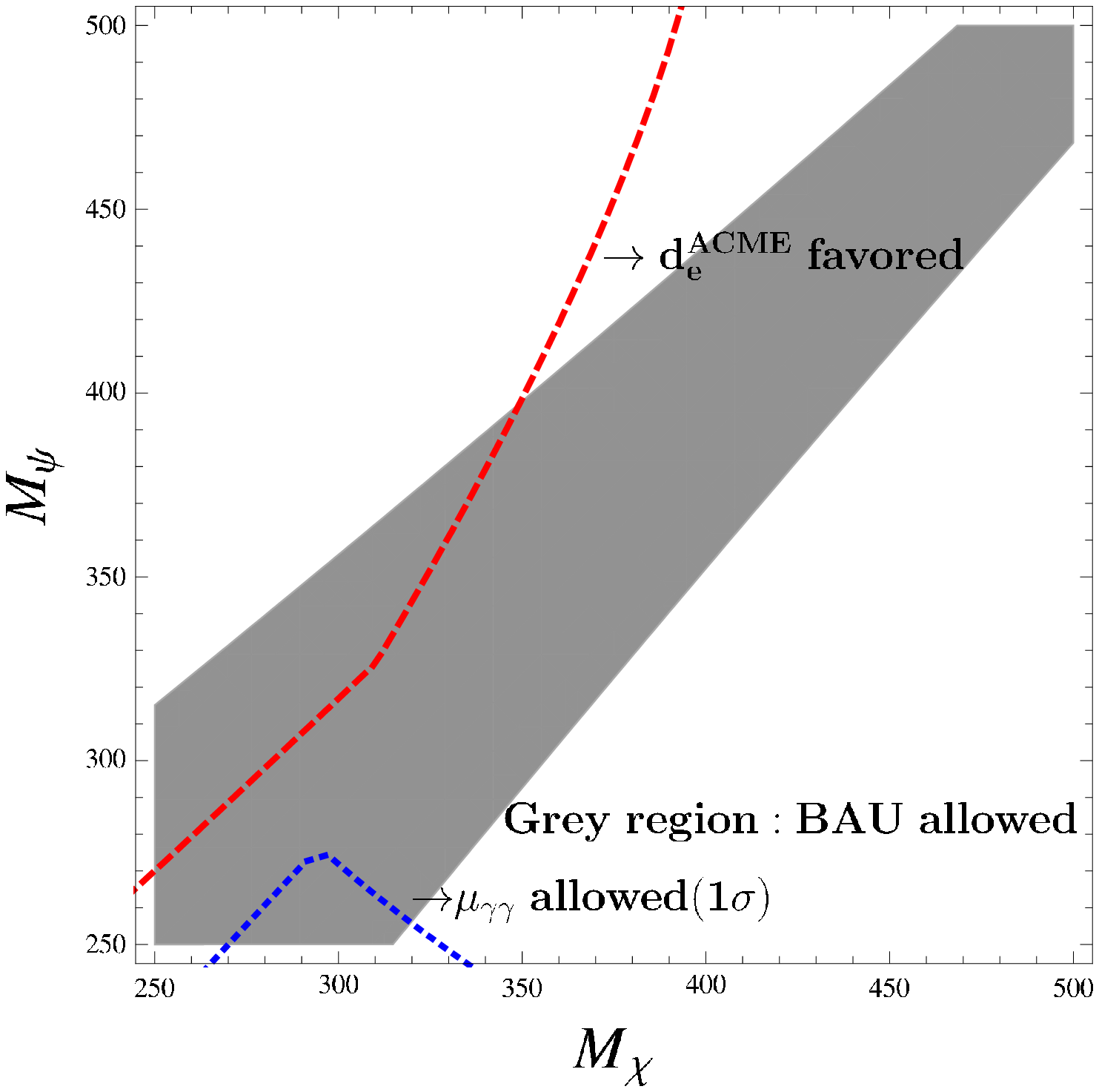}
\end{center}
\caption{Left panel: Contours of constant $Y_B\times 10^{10}$ in the $y-\delta$ plane. The input  fermion mass matrix input parameters are the same as for Fig. \ref{density}; other parameters are given in Table \ref{aaa}. Right panel:  Region consistent with observed $Y_B$ (gray region) in the $M_\chi-M_\psi$ plane. In each panels, the region to the left of the red dashed line is consistent with the ACME $ d_e$ limit at $95\% $ C.L., while the region surrounded to the left or above  the blue dotted line is consistent with the combined ATLAS and CMS  $\mu_{\gamma\gamma}$ result. We have set $|y_1|=5/3 y_1^\prime =y_2=0.6$, $y_2^\prime =0$ and $\delta =\pi/10$ generating the right panel.}\label{contour}\end{figure}

The contours of constant $n_B/s $ ( in units of $10^{-10}$) in the $y$-$\delta$ plane are indicated in the left panel of Fig.~\ref{contour}.  The initial input of the fermion mass matrix is the same as that given in the caption of Fig.~\ref{density} and other initial inputs are given in Table.~\ref{aaa}.  The region to the left of the blue dotted line satisfies the constraint of  the weighted average values of $\mu_{\gamma\gamma}$ measured by CMS and ATLAS. The region to the left of the  red dashed line obeys the constraint from the electron EDM measured obtained by the ACME experiment. We observed that the regions favored by observed Higgs diphoton rate and non-observation of the electron EDM overlap with regions of parameter space wherein a sizable portion of the baryon asymmetry is generated.  {

We also observe that both Higgs diphoton rate and charged lepton EDM depend nontrivially on  the new fermion masses. To illustrate, we plot in the right panel of Fig. \ref{contour} the region consistent with the WMAP+{ Planck} value for the observed baryon asymmetry (in gray)  in the $M_\chi-M_\psi$ plane. The region to the right of the red dashed line fulfills the constraints from the electron EDM. The region above the blue dotted line corresponds to the $\mu_{\gamma\gamma}$ $1\sigma$-allowed region.  We have assumed that $|y_1|=5/3 y_1^\prime =y_2=0.6$, $y_2^\prime =0$ and $\delta =\pi/10$ in obtaining this plot. We now comment on several features of this figure. First, since the contributions of $\chi$ and $\psi$ to the electron EDM partly cancel with each other, there is region for small $M_\chi$ satisfying the electron EDM constraint. Second, we note that the CP-violating EWBG source is resonantly enhanced when $M_\chi\approx M_\psi$; consequently, the $Y_B$-allowed region gives a diagonal band about the line of unit slope. Third, the present $\mu_{\gamma\gamma}$ constraints are not significant, as the 95\% C.L. allowed region covers the entire plane shown\footnote{Hence, we show only the $1\sigma$ line for illustrative purposes.}. Consequently, we see that there exists a substantial region of mass parameter space where the various phenomenological constraints are fulfilled. That being said, a factor of two improvement in precision on $\mu_{\gamma\gamma}$ would convert the present $1\sigma$ line roughly into a 95\% C.L. bound, indicating the possibility that a study of the diphoton rate might probe a region of the EWBG-viable parameter space not presently excluded by the electron EDM.

Looking further to the future, it is instructive to consider the prospective  parameter space sensitivity  from the next generation EDM experiments and future precision measurements of Higgs-diphoton rate.   To that end, we plot in the left-panel of the Fig. \ref{contour1} the contours of $Y_B\times 10^{10}$ in the $y-\delta$ plane, where $d_e <10^{-1}\times d_e^{\rm ACME}$ for the region to the right of the red-dashed line, while $\mu_{\gamma \gamma}-1$ is no larger than $2\%$  and $10\%$  at 95\% confidence for the regions to the left of  black-dashed line and green-dot-dashed lines, respectively. Should both measurements achieve an order of magnitude improvement in sensitivity, then $d_e$ would continue to probe most of the indicated parameter space except for small $y$ or small $\delta$, with $\mu_{\gamma\gamma}$ providing some sensitivity for the latter. Moreover, a reduction in the  $d_e$ bound by a factor of ten would preclude achieving the observed BAU for the values of mass parameters assumed in this panel. On the other hand, suppression of $d_e$ (again, possibly due to Higgs-singlet mixing) would leave open a more substantial region of parameter space that could be probe by the Higgs diphoton decays.

These features are also apparent when one considers the BAU-viable region in the space of mass parameters. In particular, we show in the right panel of Fig. \ref{contour1}, the region consistent with observed $Y_B$ (gray region) in the $M_\chi-M_\psi$ plane, by setting $|y_1|=10/3 y_1^\prime =1/3 y_2=0.3$, $y_2^\prime =0$ and $\delta =\pi/10$. The change in signal strength $\delta \mu_{\gamma \gamma } \equiv \mu_{\gamma\gamma}-1 <0.02 $ for the region above the blue dashed line and $d_e < 0.1 \times d_e^{\rm ACME}$ for the region to the right (and above) the green lines, while sufficient baryon asymmetry can be generated for the region in gray.   Again, we see that the prospective electron EDM provides a considerably more powerful probe of the EWBG-viable parameter space, unless the presence of additional contributions lead to a $d_e$ suppression. Assuming the absence of the latter, a null result for $d_e$ could, nevertheless, allow small slices of the EWBG-allowed mass space of the indicated choice of CPV phase and Yukawa coupling strength.


\begin{figure}[t]
\begin{center}
\includegraphics[width=7.0cm,height=6cm,angle=0]{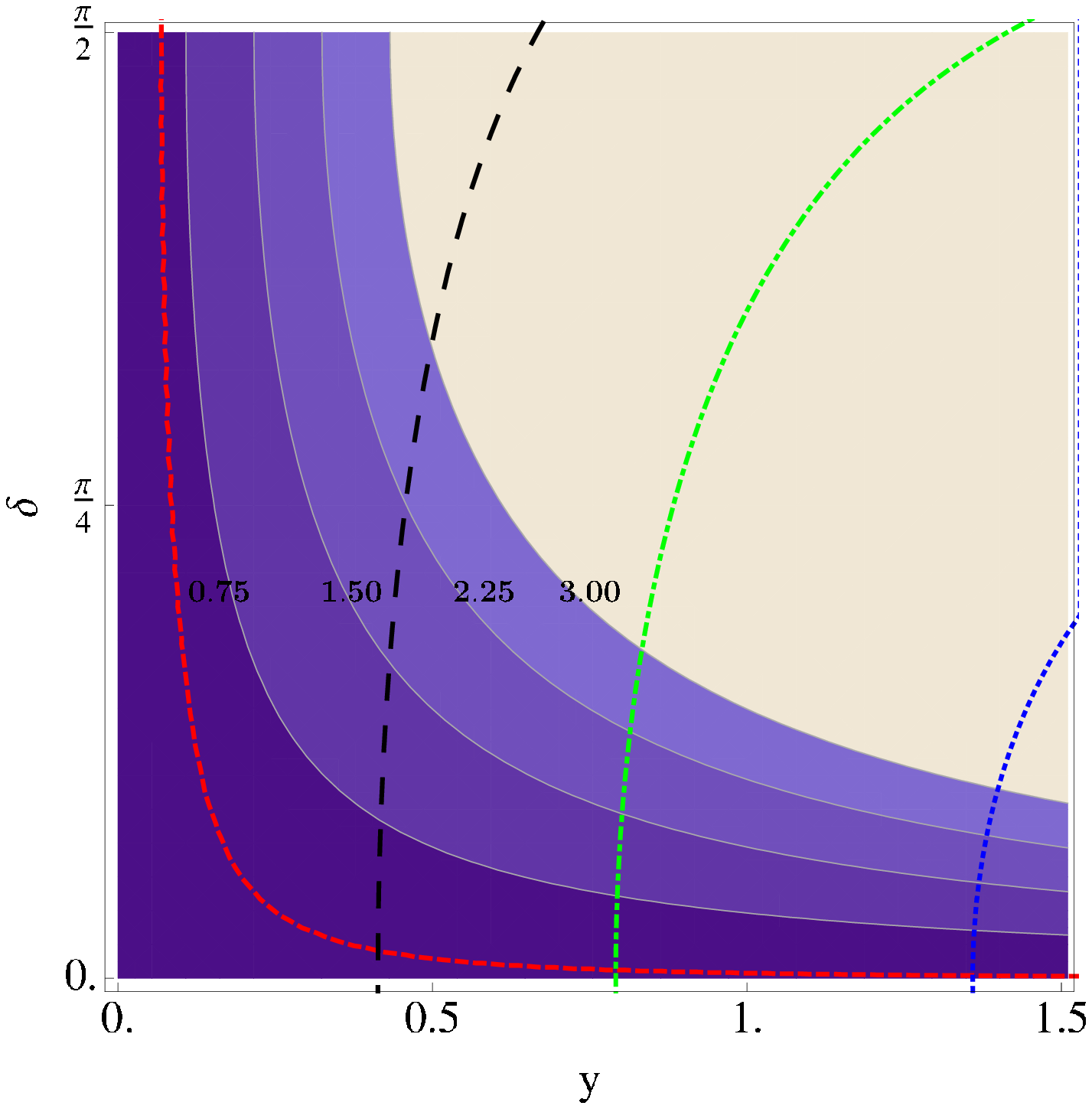}
\hspace{0.5cm}
\includegraphics[width=7.0cm,height=6cm,angle=0]{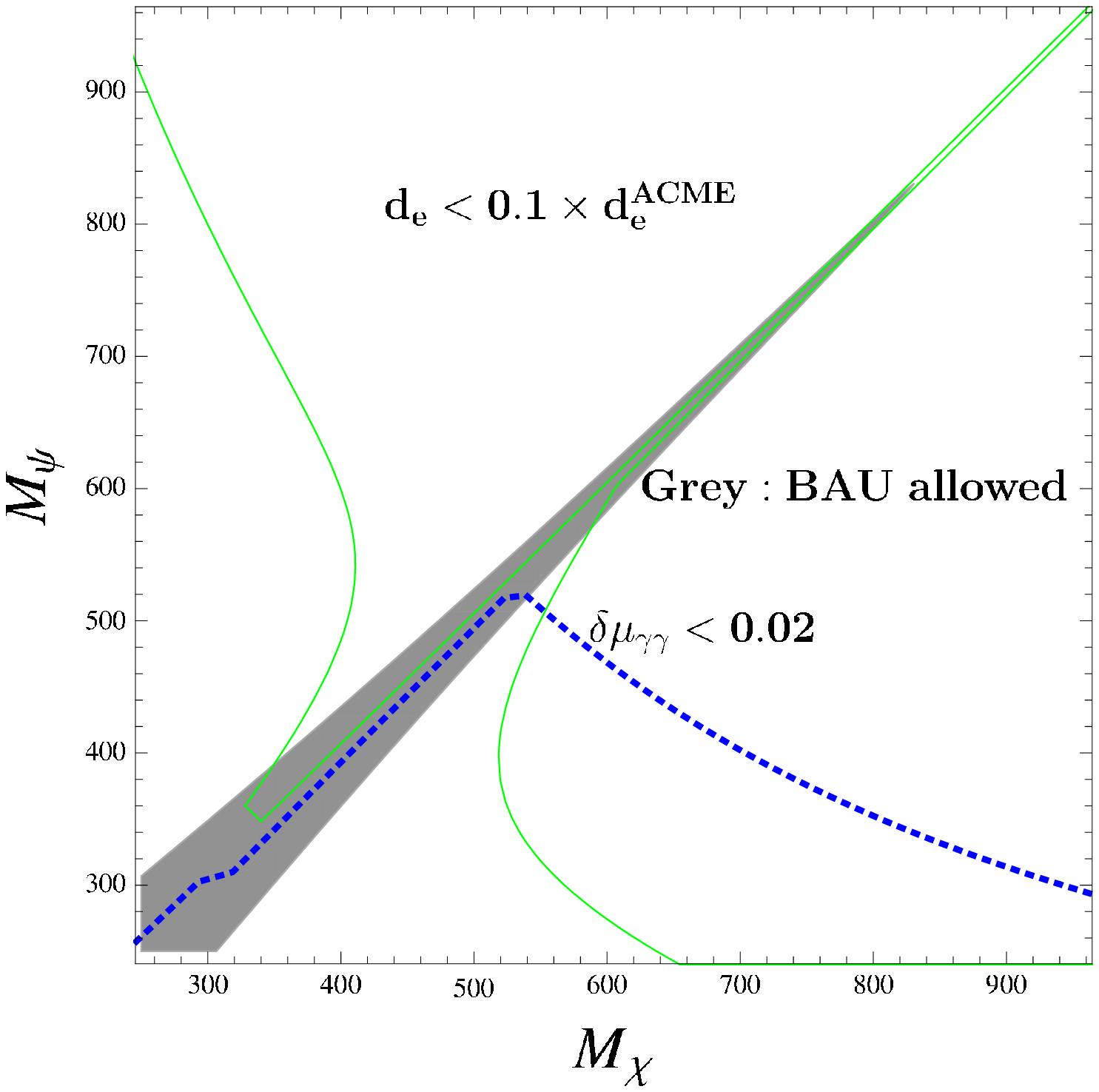}
\end{center}
\caption{Left panel: Contours of constant $Y_B\times 10^{10}$ in the $y-\delta$ plane. The input  fermion mass matrix input parameters are the same as for Fig. \ref{density}; other parameters are given in Table \ref{aaa}. The region to the left of the red dashed line is consistent with the  $0.1 \times d_e^{\rm ACME}$ limit, while the regions  to the left  the blue dotted, green dot-dashed, and black dashed lines are consistent with  $\mu_{\gamma\gamma}$ being within $20\%$, $10\%$, and $2\%$ deviation from 1, respectively. Right panel:  Region consistent with observed $Y_B$ (gray region) in the $M_\chi-M_\psi$ plane. The region outside the green line would be allowed from a $0.1 \times d_e^{\rm ACME}$ limit, while the 95\% C.L. region allowed by one percent agreement of the Higgs diphoton with the SM prediction lies above the blue dashed line.  We have set $|y_1|=10/3 y_1^\prime =1/3 y_2=0.3$, $y_2^\prime =0$ and $\delta =\pi/10$ generating the right panel.}\label{contour1}\end{figure}

\section{Concluding remarks}

Explaining the origin of the BAU remains a compelling open problem and one that may be addressed by new physics at the electroweak to TeV scale. With the discovery of the Higgs-like scalar, measurements of its properties provide new probes of such possible new interactions, in principle complementing those provided by direct searches for new scalars and low-energy, high sensitivity tests of CP invariance with EDM searches. Here, we have illustrated the interplay of these two observables by considering one of the most widely considered scalar sector extensions, the 2HDM, augmented with vector like fermions having only electroweak interactions. This scenario introduces a plethora of new CP-violating phases that may facilitate EWBG on the one hand and give rise to observable signatures in EDM searches and loop-induced Higgs decay processes on the other. Working in a restricted but illustrative region of the model parameter space\footnote{We emphasize that we have considered only a limited set of the underlying CP-violating phases and that the BAU-viable parameter space is likely to be much broader than apparent from the illustrative results given here.}, we find that it is possible for this scenario to give rise to the observed BAU while satisfying constraints from the non-observation of the electron EDM and present LHC results
for the Higgs to diphoton decay rate. The complementarity of the two experimental probes considered here is most apparent in the right panel of Figure 2, where we observe their different sensitivities to the new fermion mass spectrum. Future improvements in the sensitivities of these two sets of observables will probe more deeply into the parameter space. In general, an order of magnitude improvement in $d_e$-sensitivity would yield a considerably more comprehensive probe of the EWGB-viable parameter space considered here as compared to a factor of ten improvement in the precision of the Higgs diphoton decay rate measurement (see the right panel of Figure 3). Should additional new interactions lead to a suppression of the impact of new physics on $d_e$, future Higgs diphoton decay studies would then yield an interesting window on the EWBG mediated by new scalar-vector like fermion interactions.
More generally, the future observation of a non-zero effect in either observable could be consistent with EWBG in this scenario.


\begin{acknowledgments}
We thank W. Marciano for suggesting this study and V. Cirigliano and Y. Zhang for useful discussions and comments on the manuscript. This work was supported in part by U.S. Department of Energy contract DE-SC0011095 (MJRM and WCHAO) and in part by the National Science Foundation under Grant No. NSF PHY11-25915 (MJRM). The authors also thank the Kavli Institute for Theoretical Physics, where a portion of this work was completed.

\end{acknowledgments}

\appendix

\section{Rephasing invariants}
\label{app:ri}

Eight new phases emerge in our model, namely $\rho_\psi \equiv {\rm Arg} (M_\psi) $, $\rho_\chi \equiv {\rm Arg} (M_\chi)$, $\rho_{M_{12}^2 }^{} \equiv {\rm Arg} (M_{12}^2 )$, $\rho_{\lambda_5}^{}\equiv {\rm Arg} (\lambda_5)$ $\rho_i\equiv {\rm Arg} (y_i) $ and $\rho_i^\prime \equiv {\rm Arg} (y^\prime_i)$$(i=1,2)$.  However, not all of these phases have physical import, as some of them can be rotated way by field redefinitions:
\begin{eqnarray}
\psi_{L,R}\to  {\rm exp} (-i \phi_{\psi_{L,R}})\ \psi_{L,R}  \; ,\\
\chi_{L,R}\to  {\rm exp} (-i \phi_{\chi_{L,R}})\ \chi_{L,R} \; ,\\
H_i \to {\rm exp} (-i \phi_{H_i} )\ H_i\; .
\end{eqnarray}
The phases get shifted to
\begin{eqnarray}
&\rho_{\psi~~~} \to & \tilde{\rho}_\psi = \rho_\psi -\phi_{\psi_L } + \phi _{\psi_R } \; ,\\
&\rho_{\chi~~~}  \to & \tilde{\rho}_\chi = \rho_\psi -\phi_{\chi_L } + \phi _{\chi_R }\; ,\\
&\rho_{i~~~} \to &\tilde{\rho}_i = \rho_i-\phi_{\psi_L} + \phi_{\chi_R}+\phi_{H_i}\; ,\\
&\rho_{i~~~}^\prime  \to &\tilde{\rho}_i^\prime = \rho_i^\prime-\phi_{\chi_L} + \phi_{\psi_R}-\phi_{H_i}\; ,\\
&\rho_{M_{12}^2}^{} \to& \tilde{\rho}_{M_{12}^2 }= \rho_{M_{12}^2 }- \phi_{H_1} + \phi_{H_2} \; ,\\
&\rho_{\lambda_5~~}^{} \to & \tilde{\rho}_{\lambda_5} =\rho_{\lambda_5}- 2\phi_{H_1} + 2\phi_{H_2} \; .
\end{eqnarray}
Clearly, not all phases in Eqs. ($A4\sim A9$) are independent.  Among our eight original phases, only four are physical. The following combinations are invariant combinations under the foregoing field redefinitions:    
\begin{eqnarray}
\phi_{i~} &\equiv& \rho_i + \rho_i^\prime  - \rho_\psi - \rho_\chi  \; ,\\
\phi_m &\equiv& \rho_1 -\rho_{2}+ \rho_{M_{12}^2 } \; , \\
\phi_V&\equiv& \rho_{\lambda_5} - 2 \rho_{M_{12}^2} \; .
\end{eqnarray}
In summary, the four independent rephasing invariants are ${\rm Arg} (y_1^{}  y_1^\prime  M_\psi^* M_\chi^*)$, ${\rm Arg} (y_2^{}  y_2^\prime  M_\psi^* M_\chi^*)$, ${\rm Arg} (y_1^{}  y_2^*  M_{12}^2)$ and ${\rm Arg} (\lambda_5^{} M_{12}^{4*})$.  

{  For the rephasing invariants defined in Section II, we have
\begin{eqnarray}
\theta_{1}+\theta_{2} &=&\theta_3 + \theta_4 \; , \label{aa}\\
\theta_5+\theta_6 &=& \theta_1 -\theta_2\; , \label{bb}\\
\theta_1 -\theta_3 &=&\theta_6 \; , \label{cc} \\
\theta_3-\theta_2 &=&\theta_5 \; , \label{dd}
\end{eqnarray}
of which only three equations are independent. For example, one may take the rephasing invariants in this model  to be  $\theta_{1,2,5,7}$, with all the other rephasing invariants being constructed from these  four invariants. 

It is useful to show that ${\cal A}$, ${\cal B}$, $|{\cal R}|$ and $|{\cal Q}|$ are rephasing invariant.  A direct calculation gives
\begin{eqnarray}
{\cal A }& = & |y_1 v_1 |^2  \times \left| 1 + \left|{y_2 v_2 \over y_1 v_1}\right| e^{-i \theta_5}\right|^2  \; , \\
{\cal B } &= & |y_1^\prime v_1 |^2 \times \left| 1 + \left|{y_2^\prime v_2 \over y_1^\prime v_1}\right| e^{-i \theta_6}\right|^2  \; , \\
|{\cal R}| &=& |M_\psi y_1^\prime v_1| \left| 1+ \left| {y_2 ^\prime v_2 \over y_1^\prime v_1 } \right| e^{i \theta_6} + \left|{y_1 M_\chi \over  y_1^\prime M_\psi} \right|e^{i \theta_1}+  \left|{y_2 v_2 M_\chi \over  y_1^\prime v_1 M_\psi} \right|e^{i\theta_4}\right| \; , \label{rrr} \\
|{\cal Q}| &=&  |M_\chi y_1^\prime v_1| \left| 1+ \left| {y_2 ^\prime v_2 \over y_1^\prime v_1 } \right| e^{i \theta_6} + \left|{y_1 M_\psi \over  y_1^\prime M_\chi} \right|e^{i \theta_1}+  \left|{y_2 v_2 M_\psi \over  y_1^\prime v_1 M_\chi} \right|e^{i\theta_4}\right| \; . \label{qqq}
\end{eqnarray}
As a result, $\theta_L $ and $\theta_R$, which only depend on  ${\cal A}$, ${\cal B}$, $|{\cal R}|$ and $|{\cal Q}|$, are rephasing invariant.  In contrast, $\delta_L $ and $\delta_R$ are not rephasing invariant, because ${\cal R}$ and ${\cal Q}$ are not rephasing invariant, as can be seen from Eqs. (\ref{rrr}) and (\ref{qqq} ).

The Yukawa couplings of the charged fermions to the SM Higg  can also be written in terms of rephasing invariants:
\begin{eqnarray}
\eta_\psi  &=&+ {y_1 \over \sqrt{2}}  \cos\theta_L \sin\theta_R e^{i(\delta_R+ \phi_{\psi_L } -\phi_{\psi_R })} +{y_1^\prime \over \sqrt{2}} \cos\theta_R \sin\theta_L e^{-i(\delta_L-\phi_{\psi_L } +\phi_{\psi_R })}  \\
&\approx & + { |y_1| \over \sqrt{2} }  c_L s_R  {\rm exp} [ i  \{{\rm Arg}(M_\psi^{} M_\psi^*( y_1 y_1^* v_1 +y_1  y_2^* v_2 )) +M_\chi^{*} M_\psi^*  ( y_1 y_1^{\prime} v_1 + y_1 y_2^{\prime  }v_2 )   \} ]  \nonumber \\
&&+ { |y_1^{\prime }| \over \sqrt{2}} c_R s_L  {\rm exp } [ i\{ {\rm Arg} ( M_\psi  M_\psi^*  (y_1^\prime y_1^{\prime * } v_1 + y_1^\prime y_2^{ \prime *} v_2 ) + M_\chi^* M_\psi^* ( y_1 y_1^\prime v_1 + y_1^\prime y_2) )\}] \nonumber \\
&=&+ { |y_1| \over \sqrt{2} }  c_L s_R  {\rm exp} \left\{ i {\rm Arg} \left[ |y_1|^2 +\left| {y_1 y_2 v_2 \over v_1}\right| e^{i\theta_5} +\left |{M_\chi y_1 y_1^\prime \over M_\psi }\right|e^{i \theta_1 } +\left |{M_\chi y_1 y_2^\prime v_2  \over M_\psi  v_1}\right| e^{i\theta_3} \right]\right\} \nonumber \\
&&+{ |y_1^{\prime }| \over \sqrt{2}} c_R s_L  {\rm exp } \left\{ i {\rm Arg} \left[ |y_1^\prime|^2 +\left| {y_1^\prime y_2^\prime v_2 \over v_1}\right| e^{i\theta_6} +\left |{M_\chi y_1 y_1^\prime \over M_\psi }\right|e^{i \theta_1 } +\left |{M_\chi y_1 y_2^\prime v_2  \over M_\psi  v_1}\right| e^{i\theta_4}  \right]\right\} \nonumber\\
\eta_\chi &=& -{y_1 \over \sqrt{2}} \cos\theta_R \sin\theta_L e^{i(\delta_L+ \phi_{\chi_L } -\phi_{\chi_R })} -{y_1^\prime \over \sqrt{2}}  \cos\theta_L \sin\theta_R e^{-i(\delta_R- \phi_{\chi_L } +\phi_{\chi_R })}  \\
&\approx & -  { |y_1 | \over \sqrt{2}}  c_R s_L {\rm exp } [i  {\rm Arg}(M_\chi^* M_\psi^* (y_1 y_1^{\prime} v_1 +y_1 y_2^\prime v_2  ) + M_\chi M_\chi^* ( y_1 y_1^* v_1 + y_1 y_2^* v_2 )) ] \nonumber \\
&&- { |y_1^\prime | \over \sqrt{2} }  c_L s_R \exp [ i {\rm Arg} ( M_\chi^* M_\psi^* (y_1^\prime y_1 v_1 + y_1^\prime y_2 v_2 ) +M_\chi M_\chi^* ( y_1^\prime y_1^{\prime*} v_1 + y_1^\prime y_2^{\prime *} v_2 )) ]  \nonumber \\
&=& -  { |y_1 | \over \sqrt{2}}  c_R s_L {\rm exp } \left\{  i {\rm Arg} \left[  |y_1|^2 +\left|{y_1 y_2 v_2 \over v_1} \right|e^{i\theta_5}+ \left| { M_\psi y_1 y_1^\prime \over M_\chi } \right| e^{i \theta_1}+ \left| {M_\psi y_1 y_2^\prime v_2 \over M_\chi v_1}\right|e^{i\theta_3}\right] \right\} \nonumber \\
&&-{ |y_1^\prime | \over \sqrt{2} }  c_L s_R \exp \left\{ i {\rm Arg} \left[  |y_1^\prime|^2 + \left| {y_1^\prime y_2^\prime v_2 \over v_1}\right| e^{i\theta_6} +\left |{M_\psi y_1 y_1^\prime \over M_\chi }\right|e^{i \theta_1 } +\left |{M_\psi y_1 y_2^\prime v_2  \over M_\chi  v_1}\right| e^{i\theta_4}  \right]\right\} \nonumber
\end{eqnarray} 
which are of course rephasing invariant.
}

Finally, we prove that $\delta_R -\delta_L + {\rm Arg} (M_\chi) - {\rm Arg} (M_\psi)$  is rephasing invariant:
\begin{eqnarray}
&&\delta_R -\delta_L + {\rm Arg} (M_\chi) -{\rm Arg}(M_\psi) \nonumber \\
=&&+  {\rm Arg} [M_\psi M_\chi (y_1^{\prime *} v_1 + y_2^{\prime *  } v_2) + |M_\chi|^2 ( y_1v_1 + y_2 v_2 ) ] \nonumber \\
&&-{\rm Arg } [ M_\psi M_\chi (y_1^{\prime * } v_1 + y_2^{\prime * } v_2 ) + |M_\psi|^2 (y_1 v_1 + y_2 v_2 )]  \nonumber \\
=&& + {\rm Arg} [M_\psi M_\chi (y_1^{\prime *} v_1 + y_2^{\prime *  } v_2) ( y_1^*v_1 + y_2^*  v_2 )+ |M_\chi|^2 |y_1v_1 + y_2 v_2 |^2  ]  \nonumber \\&&-{\rm Arg } [ M_\psi M_\chi (y_1^{\prime * } v_1 + y_2^{\prime * } v_2 ) (y_1^* v_1 + y_2^* v_2 )+ |M_\psi|^2 |y_1 v_1 + y_2 v_2 |^2 ]  \nonumber \\
=&& +{\rm Arg} [M_\psi^*  M_\chi^* y_1 y_1^\prime v_1^2 + M_\psi^*  M_\chi^* y_1 y_2^\prime v_1v_2 + M_\psi^*  M_\chi^* y_2 y_1^\prime v_1v_2 + M_\psi^*  M_\chi^* y_2 y_2^\prime v_2^2 \nonumber \\
&& +  |M_\psi|^2 |y_1 v_1 + y_2 v_2 |^2   ] \nonumber \\
&& -{\rm Arg} [M_\psi^*  M_\chi^* y_1 y_1^\prime v_1^2 + M_\psi^*  M_\chi^* y_1 y_2^\prime v_1v_2 + M_\psi^*  M_\chi^* y_2 y_1^\prime v_1v_2 + M_\psi^*  M_\chi^* y_2 y_2^\prime v_2^2 \nonumber \\
&& +  |M_\chi|^2 |y_1 v_1 + y_2 v_2 |^2   ] \label{proving}
\end{eqnarray}
Clearly, Eq.~(\ref{proving}) is rephasing invariant, because they are written as rephasing invariants that defined above.

\end{document}